\documentclass[12pt]{article}

\usepackage[T1]{fontenc}
\usepackage[utf8]{inputenc}
\usepackage{authblk}

\usepackage{amsfonts,amsmath,color}
\usepackage[all]{xy}

\usepackage{amssymb}
\usepackage{amstext}
\usepackage{latexsym}
\usepackage{graphicx}
\usepackage{epsf}
\usepackage{epsfig}
\usepackage{subfigure}
\usepackage{graphics}
\usepackage{verbatim}
\usepackage{color}
\usepackage{hyperref}

\def\C{\ensuremath{\mathbb C}}
\def\R{\ensuremath{\mathbb R}}

\newcommand{\be}{\begin{equation}}
\newcommand{\ee}{\end{equation}}
\newcommand{\bea}{\begin{eqnarray}}
\newcommand{\eea}{\end{eqnarray}}

\renewcommand{\c}[1]{\ensuremath{\overline{#1}}}

\newcommand{\da}{^{\dagger}}
\newcommand{\vac}{\ensuremath{\left|0,0\right\rangle}}
\newcommand{\pa}[2]{\frac{\partial #1}{\partial #2}}

\newcommand{\ma}[4]{\left[\begin{array}{cc}#1&#2\\#3&#4\end{array}\right]}

\newcommand{\blue}[1]{\textcolor[rgb]{0, 0, 0.9}{#1}}

\newcommand{\z}{\ensuremath{(z,\overline{z})}}

\newcommand{\an}[1]{a^{\alpha}_{#1}{}^{\dagger}} 
\newcommand{\ao}[1]{a^{\alpha}_{#1}}
\newcommand{\ja}[1]{J^{\alpha}_{#1}}

\newcommand{\fr}[1]{\mathfrak{#1}}

\newcommand{\en}{\end{equation}}

\newtheorem{thm}{Theorem}
\newtheorem{cor}[thm]{Corollary}

\newtheorem{defi}{Definition}[section]
\newtheorem{lem}[defi]{Lemma}

\newtheorem{Theo}{Theorem}[section]

\newcommand{\bedefin}{\begin{defi}}
\newcommand{\findefi}{\end{defi} \medskip}

\newcommand{\betheo}{\begin{theorem}$\!\!${\bf \,\,\,}}
\newcommand{\entheo}{\end{theorem}}
\newcommand{\enth}{\end{theorem}}

\newcommand{\becor}{\begin{cor}$\!\!${\bf .}}
\newcommand{\encor}{\end{cor}}

\newcommand{\belem}{\begin{lem}$\!\!${\bf }}
\newcommand{\enlem}{\end{lem}}

\newcommand{\prf}{\noindent{\bf{\small Proof.}\,\,\,\,}}
\newcommand{\qed}{\hfill $\blacksquare$}

\newcommand{\ena}{\end{eqnarray}}

\newcommand{\beano}{\begin{eqnarray*}}
\newcommand{\enano}{\end{eqnarray*}}

\newcommand{\bee}{\begin{enumerate}}
\newcommand{\ene}{\end{enumerate}}

\newcommand{\bei}{\begin{itemize}}
\newcommand{\eni}{\end{itemize}}

\newcommand{\betab}{\begin{tabular}}
\newcommand{\entab}{\end{tabular}}

\newcommand{\bd}{\begin{displaymath}}

\newcommand{\h}{{\mathfrak H}}

\title{Some Biorthogonal Families of Polynomials Arising in Noncommutative Quantum Mechanics}
\author[1]{F. Balogh}
\author[2,3]{Nurisya M. Shah}
\author[4]{S. Twareque Ali}

\affil[1]{SISSA, Trieste, Italy}
\affil[2]{Department of Physics,
Concordia University, Montr\'eal, Qu\'ebec, Canada}
\affil[3]{Department of Physics, Faculty of Science, Universiti Putra Malaysia, Serdang, Malaysia}
\affil[4]{Department of Mathematics and Statistics, Concordia University, Montr\'eal, 
Qu\'ebec, Canada}

\begin{document}

\maketitle
\begin{abstract}
In this paper we study families of complex Hermite polynomials and construct deformed versions of them, using a $GL(2, \mathbb C)$ transformation. This construction leads to the emergence of biorthogonal families of deformed complex Hermite polynomials, which we then study in the context of a two-dimensional model of noncommutative quantum mechanics.
\end{abstract}

\section{Introduction}\label{sec-intro}
  Noncommutative quantum mechanics is a highly active area of current research.
The motivating factor here is the belief that a  modification of standard quantum mechanics
is needed to model physical space-time at very short distances. One way to
introduce such a modification is to alter the the canonical commutation relations of
quantum mechanics. One can then study, for example, the effect of such a modification on well-known Hamiltonians, such as the harmonic oscillator or the Landau problem and their energy
spectra. For some recent work in this direction see, e.g.,
\cite{bengeloun1,bengeloun,gamboa,jing,kang,muthu,scho}. In this paper we work with a similar version of noncommutative quantum mechanics, i.e., one describing a system with two degrees of freedom. However, our aim here is to study certain
associated classes of complex biorthogonal polynomials, arising as a consequence of the altered commutation relations.  These biorthogonal polynomials appearing in much the same way as
the {\em complex Hermite polynomials} \cite{ghanmi08,intint,ito,matsu,mcgee}, arise in the standard quantum mechanics of a system with two degrees of freedom. 

The main body of our results on the biorthogonal complex Hermite polynomials is presented in Sections \ref{sec-def-herm} -- \ref{sec-biorth-poly}, in particular in Theorem \ref{theor-biorth}. The relationship of these polynomials to noncommutative quantum mechanics is discussed in Section \ref{sec-ncqm}.

We start with the usual quantum mechanical commutation relations
\be
  [Q_i , P_j ] = i \delta_{ij}I,\quad  i, j = 1,2\, .
\label{nc-comm-relns1}
\en
Here the $Q_i, P_j$ are the quantum mechanical position and momentum observables, respectively.
In   {non-commutative quantum mechanics} one imposes the additional commutation relation
\be
[Q_1, Q_2] = i\vartheta I,
\label{nc-comm-relns10}
\en
where $\vartheta$ is a   {small, positive} parameter  which measures the additionally introduced
noncommutativity between the observables of the two spatial coordinates.
The limit $\vartheta = 0$ then corresponds to standard (two-dimensional) quantum mechanics.
One could also impose a second non-commutativity between the two momentum operators:
\be
  [P_1 , P_2 ] = i\gamma I\; ,
\label{non-comm-relns2}
\en
where $\gamma$ is yet another positive parameter. Physically, such a commutator would mean that there is a magnetic field in the system.

The $Q_i$ and $P_i, \; i = 1,2$, satisfying the modified commutation relations (\ref{nc-comm-relns10}) and (\ref{non-comm-relns2}) can be written in terms of the
standard quantum mechanical position and momentum operators $\hat{q}_i,  \hat{p}_i , \;
 i=1,2$, with
 $$[\hat{q}_i , \hat{p}_j ] = i\delta_{ij}, \;\; \; [ \hat{q}_i ,\hat{q}_j] =
 [ \hat{p}_i ,\hat{p}_j] = 0.$$
One possible representation is
 \bea
   Q_1 = \hat{q}_1 - \frac {\vartheta}2 \hat{p}_2 & \qquad & P_1 = c\hat{p}_1 + d\hat{q}_2
               \qquad c = \frac 12 (1 \pm \sqrt{\kappa}), \;\; d = \frac 1{\vartheta}
      (1 \mp \sqrt{\kappa})
               \nonumber\\
 Q_2 = \hat{q}_2  + \frac {\vartheta}2 \hat{p}_1 & \qquad & P_2 = c\hat{p}_2 - d\hat{q}_1
 \qquad \kappa = 1 - \gamma \vartheta, \;\;
      \gamma \neq \frac 1{\vartheta}
 \label{non-can-transf1}
 \ena

In this paper, we shall assume such a non-commutative system, however,  with the additional restriction $$\vartheta = \gamma .$$ Then, introducing the annihilation and creation operators,
\be
  A_i = \frac 1{\sqrt{2}}(Q_i + iP_i), \quad  A_i^\dag =
  \frac 1{\sqrt{2}}(Q_i - iP_i), \quad i = 1,2,
\label{crann-ops}
\en
we have the modified commutation relations,
\be
  [A_i, A_i^\dag ] =1, \quad [A_i, A_j ] = 0, \quad [A_1, A_2^\dag ] = i\vartheta ,
  \quad i,j = 1,2\; .
\label{ceann-op-comm}
\en
Note, in particular that
$$[A_1, A_2 ] =0 \quad \Longrightarrow \quad [A_1^\dag , A_2^\dag] =0, $$
which means that one still has two  independent bosons and the operators $A_1$ and $A_2$ still have a common ground state, which we may conveniently denote by $\vert 0,0\rangle$.

  It is well known that the Hermite polynomials (in a real variable) are naturally
associated to the commutation relations $[a, a^\dag] = 1$ for a single bosonic degree of freedom. Writing them as
$H_n (x), \; n =0,1,2, \ldots , \;\; x \in \mathbb R$, they satisfy the orthogonality relations:
\be
  \int_{\mathbb R} H_m (x) H_n (x)\; e^{-x^2}\; dx = \sqrt{\pi} 2^n n!\; \delta_{mn}\;,
\label{real-herm-orth}
\en
and are obtainable using the formula:
\be
H_n (x) = (-1)^n e^{x^2} \left(\frac d{dx}\right)^n e^{-x^2}\; ,
\label{hp-form1}
\en
or using the generating function,
\be
  e^{2xz - z^2} = \sum_{n=0}^\infty \frac {z^n}{n!} H_n (x)\; .
\label{real-hp-gen-fcn}
\en
On the Hilbert space
$L^2 (\mathbb R , e^{-x^2}dx),$
the operators of creation and annihilation are,
\be
  a^\dag = \frac 1{\sqrt{2}}\left(2x - \frac d{dx}\right)\; , \qquad a =
  \frac 1{\sqrt{2}}\frac d{dx}\; , \qquad [a, a^\dag ] =1\; .
\label{cr-an-ops1}
\en
On this space, the normalized Hermite polynomials $h_n$ can be obtained as:
\be
  h_n(x) := \frac 1{[{\sqrt{\pi}\; 2^n\; n!}]^{\frac 12}}\; H_n (x) \quad \text{and}
  \quad h_n =  \frac {(a^\dag)^n}{\sqrt{n!}}h_0\; , 
\label{normaliz-rhp}
\en
where $h_0$, the ground state, is the constant function,
$$ h_0 (x)  = \frac 1{\pi^{\frac 14}}, \quad x \in \mathbb R\; .$$
The vectors $h_n$ form an orthonormal basis of $L^2 (\mathbb R , e^{-x^2}dx)$. All this, of course, is standard and well-known.

A second representation of the commutation relation $[a, a^\dag] =1$, and the one that will be more pertinent to the present work, is on the Hilbert space (Fock-Bargmann space) $L^2_{\text{anal}} (\mathbb C , d\nu\z)$ of all analytic functions of a complex variable  $z = x + iy$, which are square integrable with respect to the measure
$$d\nu\z = e^{-|z|^2}\;\dfrac{dz\wedge d\c{z}}{i2\pi}= e^{-[x^2 + y^2]}\; \dfrac {dx\;dy}\pi\; .$$
On this space the creation and annihilation operators take the form $a^\dag = z$ (operator of multiplication by $z$) and $a = \partial_z$, respectively. The normalized ground state is the constant function $h_0 (z) = 1, \;\; z \in \mathbb C$. The orthonormal basis, built again as  in (\ref{normaliz-rhp}), is now given by the monomials
\be
\quad h_{n} (z) =  \frac {(a^\dag)^n}{\sqrt{n!}}h_{0} = \frac {z^n}{\sqrt{n!}}\; .
\label{anal-basis}
\en
For two independent bosons, the two sets of creation and annihilation operators $a^\dag_i, a_i, \;\; i = 1,2$, satisfy the commutation relations $[a_i , a_j^\dag] = \delta_{ij}$, and this set can be irreducibly represented on the full Hilbert space $\h(\mathbb C) = L^2 (\mathbb C , d\nu\z)$,
via the operators ~\cite{sta}
\be
a_1 = \partial_z , \;\; a_1^\dag = z - \partial_{\overline{z}}, \qquad
a_2 = \partial_{\overline{z}} , \;\; a_2^\dag = \overline{z} - \partial_z,
\label{2-bos-ops}
\en
Note that $L^2_{\text{anal}} (\mathbb C , d\nu\z)$ is a proper subspace of $\h(\mathbb C)$.
Using again the ground state $h_{0,0}\z \equiv 1$, one can build an orthonormal basis for
$\h(\mathbb C)$:
\be
h_{m,n}\z = \frac {(a_1^\dag )^m\; (a_2^\dag )^n}{\sqrt{m!\; n!}}\; h_{0,0} \; = \;
       \frac {(z - \partial_{\overline{z}})^m\; (\overline{z} - \partial_z)^n}{\sqrt{m!\; n!}}\; 1,
\label{full-basis}
\en
$m,n = 0,1,2, \ldots $.
It is clear that $h_{m,0}\z= h_m (z)$. The functions $h_{m,n}\z$ have the explicit forms (see, for example, \cite{ghanmi08}):
\be
  h_{m,n}  \z  = \sqrt{m!\;n!} \sum_{j= 0}^{m\curlyvee n}\frac {(-1)^j}{j!}
    \frac {z^{m-j}}{(m-j)!}\; \frac {\overline{z}^{n-j}}{(n-j)!}\; ,
\label{comp-herm-poly9}
\en
where $m\curlyvee n$ denotes the smaller of the two numbers $m$ and $n$. Moreover, it is
easy to verify that the functions
$H_{m,n}\z = \sqrt{m! n!}\; h_{m,n}\z$ are also obtainable as
\be
 H_{m,n} (\overline{z} , z)  = (-1)^{m+n} \;e^{\vert z\vert^2}\partial^m_z \partial^n_{\overline z}
   \; e^{-\vert z\vert^2}\; ,
\label{comp-herm-poly}
\en
a relation which should be compared to (\ref{hp-form1}). By analogy, the functions $H_{m,n}$ (of which the $h_{m,n}$ are just normalized versions) are called
{\em complex Hermite polynomials} \cite{ghanmi08,intint}.

  In the sequel we shall basically ``deform'' the relation (\ref{full-basis}) to obtain families
of {\em generalized biorthogonal Hermite polynomials}, in which the operators $a_i\; a_i^\dag, \;\; i =1,2$ will be replaced by operators similar to those in (\ref{crann-ops}) of noncommutative quantum mechanics.

 The rest of this paper is organized as follows. In Section \ref{sec-abs-prelim} we
lay down some abstract preliminaries connected with Hermite polynomials and construct generating functions, using an operator technique. In Section \ref{sec-def-herm} we introduce the deformed complex Hermite polynomials, obtain some of their immediate properties and work out the representation of $GL(2, \mathbb C)$ which gives rise to the deformed polynomials. In Section \ref{sec-biorth-poly} we introduce the families of biorthogonal deformed complex Hermite polynomials. Section \ref{sec-ncqm} is devoted to a study of the pertinence of the above results to a two-dimensional model of noncommutative quantum mechanics. Finally, in Section \ref{sec-bilin-ops} we look at some second order generators built out of the deformed creation and annihilation operators introduced earlier and identify the Lie algebras generated by them.

\section{Some abstract preliminaries and generating functions}\label{sec-abs-prelim}
 We go back to the algebra associated to two independent bosons, generated by the usual lowering and raising operators $a_1, a_2$ and $a_1\da, a_2\da$ respectively,
satisfying the commutation relations
\be
[a_i, a_j] = 0\ ,\ [a_i\da, a_j\da] = 0\ ,\ [a_i, a_j\da] = \delta_{ij}\ , \quad i,j=1,2\ .
\ee
Assuming an irreducible representation of this system in an abstract Hilbert space
$\h$, the lowering operators annihilate the vacuum state $\vac$,
\be\label{e:vac}
a_i \vac =0, \qquad i=1,2\; .
\ee
The Hilbert space $\h$ is then spanned by the orthonormal basis set,
\be
\left|k,l\right\rangle =\frac{1}{\sqrt{k!l!}}(a_1\da)^{k}(a_2\da)^{l}\vac\ , \quad k,l =0,1,2,\dots\ ,
\ee
with
\begin{align*}
a_1\da \left|k,l\right\rangle &= \sqrt{k+1}\left|k+1,l\right\rangle\ ,\ a_2\da \left|k,l\right\rangle = \sqrt{l+1}\left|k,l+1\right\rangle\ .
\end{align*}
The vector-valued function
\be
F(u, \c{u}) = \sum_{k,l=0}^{\infty}\frac{u^k\c{u}^l}{\sqrt{k!\;l!}}\left|k,l\right\rangle
\ee
will serve as a useful book-keeping device in the subsequent calculations based on the identities
\be
\pa{F}{u}(u,\c{u}) = a_1\da F(u,\c{u}) \mbox{ and }\pa{F}{\c{u}}(u,\c{u}) = a_2\da F(u,\c{u})\ .
\ee
Obviously,
\be
F(u,\c{u}) = e^{ua_1\da+\c{u}a_2\da}\vac\ .
\ee

On the Hilbert space
$\h(\C) = L^2(\C, d\nu\z)$, introduced above,
\be
\vac \mapsto h_{0,0}\z= 1,\qquad \text{for all} \quad z,\c{z},
\ee
and for any vector $f \in \h(\C)$
\begin{align*}
[e^{ua_1\da}f](z,\c{z}) &= e^{uz}f(z,\c{z}-u)\\
[e^{\c{u}a_2\da}f](z,\c{z}) &= e^{\c{u}\c{z}}f(z-\c{u},\c{z})\\
\end{align*}
and therefore
\be
F(u,\c{u}) = e^{ua_1\da+\c{u}a_2\da}\vac = e^{uz+\c{u}\c{z} -u\c{u}}\ .
\ee
Expanding this with respect to $u$ and  $\c{u}$  and comparing with (\ref{full-basis}) gives the {\em generating function} for the complex Hermite polynomials in (\ref{comp-herm-poly9}) - (\ref{comp-herm-poly}),
\be
e^{uz+\c{u}\c{z} -u\c{u}} = \sum_{k,l=0}^{\infty}h_{k,l}(z,\c{z})\frac{u^k\c{u}^l}{\sqrt{k!\;l!}}
= \sum_{k,l=0}^{\infty}H_{k,l}(z,\c{z})\frac{u^k\c{u}^l}{k!\;l!}\; .
\label{comp-gen-fcn}
\ee

Consider another well-known representation of the real Hermite polynomials~\cite{mes} on the Hilbert space
$\h = L^2(\R^2, dx_1dx_2)$, with
\be
\vac \mapsto h_{0,0}(x_{1},x_{2}) = e^{-\frac{1}{2}(x_1^2+x_2^2)}
\ee
and
\begin{align*}
[a_i\da f](x_1,x_2)&= \frac{1}{\sqrt{2}}\left(x_{i} -\pa{}{x_{i}}\right)f(x_1,x_2)\quad i=1,2\ ,\\
[a_i f](x_1,x_2)&= \frac{1}{\sqrt{2}}\left(x_{i} +\pa{}{x_i}\right)f(x_1,x_2)\quad i=1,2\ .
\end{align*}

Then, by the Baker-Campbell-Hausdorff formula,
\begin{align*}
[e^{ua_1\da}f](x_1,x_2) &= e^{ux_1-\frac{1}{2}u^2}f(x_1-u,x_2)\\
[e^{\c{u}a_2\da}f](x_1,x_2) &= e^{\c{u}x_2-\frac{1}{2}\c{u}^2}f(x_1,x_2-\c{u})\\
\end{align*}
and therefore
\begin{align*}
F(u,\c{u}) &= e^{ua_1\da+\c{u}a_2\da}\vac\\
 &= e^{ux_1-\frac{1}{2}u^2+\c{u}x_2-\frac{1}{2}\c{u}^2-\frac{1}{2}(x_1-u)^2-\frac{1}{2}(x_2-\c{u})^2}\\
&= e^{2ux_1-u^2+2\c{u}x_2-\c{u}^2-\frac{1}{2}x_1^2-\frac{1}{2}x_2^2}\ .
\end{align*}
Expanding this with respect to $u$ and $\c{u}$ and comparing with (\ref{real-hp-gen-fcn}) gives the generating function for products of real \emph{Hermite polynomials} in two-variables~\cite{mes}
\be
F(u,\c{u}) = e^{-\frac{1}{2}[x_1^2 + x_2^2]}\sum_{k,l=0}^{\infty} \frac{u^k\c{u}^l}{k!\;l!}H_{k}(x_1)\;H_l(x_2)\ .
\ee

\section{Deformed generalized Hermite polynomials}\label{sec-def-herm}
Going back to (\ref{full-basis}), we now {\em deform} the complex Hermite polynomials $h_{m,n}$ essentially by replacing the operators $a_1^\dag$ and $a_2^\dag$ by linear combinations of these. Specifically, we define the operators
\bea\label{e:lin}
{a^{g}_1}\da = g_{11}a_1\da+g_{21}a_2\da\ , &\quad& {a^{g}_2}\da = g_{12}a_1\da+g_{22}a_2\da\nonumber\\
{a^{g}_1} = \overline{g_{11}}a_1 + \overline{g_{21}} a_2\da\ , &\quad& a^{g}_2 = \overline{g_{12}} a_1 + \overline{g_{22}}a_2\; .
\ena
parametrized by a $2\times 2$ invertible complex matrix
\be\label{e:gmat}
g= \ma{g_{11}}{g_{12}}{g_{21}}{g_{22}} \in GL(2,\C)\ .
\ee
The {\em $g$-deformed basis elements} are then defined to be
\be
|k,l\rangle_g = \frac{1}{\sqrt{k!l!}}({a_1^{g}}\da)^k({a_2^{g}}\da)^l\vac \qquad k,l = 0,1,\dots
\ee
The generating function of the $g$-deformed basis is given by
\be
F_{g}(u,\c{u}) = \sum_{k,l=0}^{\infty}\frac{u^k\c{u}^l}{k!\;l!}({a^{g}_1}\da)^k({a^{g}_2}\da)^l\vac
=e^{u{a^{g}_1}\da+\c{u}{a^{g}_2}\da}\vac
\ee
which can be written as
\begin{align*}
F_{g}(u,\c{u}) &= e^{u{a^{g}_1}\da+\c{u}{a^{g}_2}\da}\vac\\
&= e^{(g_{11}u+g_{12}\c{u})a_1\da+(g_{21}u+g_{22}\c{u})a_2\da}\vac\\
&=F(g_{11}u+g_{12}\c{u},g_{21}u+g_{22}\c{u})\ .
\end{align*}
In particular, we get the analogue of (\ref{comp-gen-fcn})
\begin{align}\label{def-gen-fcn}
F_{g}(u,\c{u}) &= \exp\left((g_{11}u+g_{12}\c{u})z+(g_{21}u+g_{22}\c{u})\c{z} -(g_{11}u+g_{12}\c{u})(g_{21}u+g_{22}\c{u})\right)\nonumber\\
&=\sum_{k,l=0}^{\infty}h_{k,l}^{g}(z,\c{z})\frac{u^k\c{u}^l}{\sqrt{k!\;l!}}
=\sum_{k,l=0}^{\infty}H_{k,l}^{g}(z,\c{z})\frac{u^k\c{u}^l}{k!\;l!}
\,
\end{align}
where the polynomials
\be\label{e:defher}
h_{k,l}^{g}(z,\c{z}) :=
 \frac {({a^g_1}^\dag)^k\; ({a^g_2}^\dag )^l}{\sqrt{k!\; l!}}\; h_{0,0}\z
 = \frac {H^g_{k,l}\z}{\sqrt{k!\;l!}}\; ,
\qquad k,l=0,1,\dots
\ee
are the {\em $g$-deformed complex Hermite polynomials}.

Our aim is to describe the operator $T_g$ defined by
\be\label{e:matT}
T_g |k,l\rangle = |k,l\rangle_g \qquad k,l=0,1,\dots
\ee
in terms of the group element $g \in GL(2,\C)$.
Consider the map
\be
P \colon \h \to \C[s,t]\qquad | k,l \rangle \mapsto \frac{1}{\sqrt{k!l!}}s^k t^l\ ,
\ee
where $\C[s,t]$ denotes the set of all complex  polynomials in the two variables $s$ and $t$.
Then
\be
P a_1\da = M_s P\ , \qquad P a_2\da = M_t P\ ,
\ee
where $M_s$ and $M_t$ stand for the operators of multiplication by $s$ and $t$ respectively.
Therefore
\be
T_{g}|k,l\rangle_g = \frac{1}{\sqrt{k!l!}}(g_{11}s+g_{21}t)^k(g_{12}s+g_{22}t)^l.
\ee
Let
\be
R_g\ \colon \ \C[s,t] \to \C[s,t] \qquad [R_g f](s,t) = f(g_{11}s+g_{21}t,g_{12}s+g_{22}t).
\ee
Then the following intertwining relation holds:
\be
PT_g = R_gP\ .
\ee
To summarize\blue{,} we have a commutative diagram:
\be\xymatrix{
H{\ar[r]^{P}\ar[d]^{T_g}}&{\C[s,t] }{\ar[d]^{R_g}}\\
H{\ar[r]^{P}}&{\C[s,t] }\\
}
\ee
The operators $R_g$ realize a  representation of $GL(2,\C)$ on the space $\C[s,t]$ and it splits into an infinite direct sum of irreducible representations
\be
\C[s,t] = \bigoplus_{L=0}^{\infty}\C_L[s,t],
\ee
where $\C_L[s,t]$ stands for the subspace of homogeneous polynomials of degree $L$:
\be
\C_L[s,t] = \mbox{span}\{s^k t^{L-k}\colon k=0,\dots, L\}\ .
\ee
If we take
\be
V=\C_1[s,t] = \mbox{span}\{s,t\},
\ee
then we see that $R_g|_{V}$ is the standard representation of $GL(2,\C)$ and that
\be
\C_L[s,t] \simeq \mbox{Sym}^L V\
\ee
($L$\blue{-}fold symmetric tensor product of $V$). This representation is irreducible.
A straightforward calculation gives
\begin{align*}
&\ \ (g_{11}s+g_{21}t)^k(g_{12}s+g_{22}t)^l\\
&= \sum_{i=0}^{k}\sum_{j=0}^{l}\binom{k}{i}\binom{l}{j}g_{11}^ig_{21}^{k-i}g_{12}^jg_{22}^{l-j}s^{i+j}t^{k+l-i-j}\\
&=\sum_{r=0}^{k+l}\left[\sum_{q=\max\{0,r-l\}}^{\min\{r,k\}}\binom{k}{q}\binom{l}{r-q}g_{11}^{q}g_{21}^{k-q}g_{12}^{r-q}g_{22}^{l+q-r}\right]s^{r}t^{k+l-r}\ .
\end{align*}
where we used the substitution $r=i+j$ and $q=i$. If we choose the basis in $\C_L[s,t]$ as
\be
f_k(s,t) = s^k t^{L-k}\qquad k=0,1,\dots, L
\ee
then
\be
R_gf_k = \sum_{r=0}^{L}\left[\sum_{q=\max\{0,r+k-L\}}^{\min\{r,k\}}\binom{k}{q}\binom{L-k}{r-q}g_{11}^{q}g_{21}^{k-q}g_{12}^{r-q}g_{22}^{L-k+q-r}\right]f_r\ .
\ee
So the matrix $M(g,L)$ of $R_g|_{C_L[s,t]}$ in the basis $\{f_k\}_{k=0}^{L}$ is given by the matrix elements
\be
M(g,L)_{rk} = \sum_{q=\max\{0,r+k-L\}}^{\min\{r,k\}}\binom{k}{q}\binom{L-k}{r-q}g_{11}^{q}g_{21}^{k-q}g_{12}^{r-q}g_{22}^{L-k+q-r} \quad 0\leq r,k \leq L\ .
\label{irrep-mat-elem}
\ee
A useful relation that we read off from the above is that
\be
   M(g,L)^* = M(g^*, L),
\label{adj-rep}
\en
the star denoting the adjoint matrix in each case.

If furthermore,  $\lambda_{1},\lambda_{2}$ are non-zero eigenvalues of $M(g,2)$, corresponding to  non-zero eigenvectors $f_{1},f_{2}$, it is possible to show that the eigenvalues of $M(g,L)$ are
\be\label{e:eigenv}
\Lambda_{L}= \{\lambda_{1}^{k} \lambda_{2}^{L-k}\, ; \, k=0,1,\ldots, L\} .
\ee
This is another useful result since (\ref{e:eigenv}) provides  complete information regarding any particular choice of roots, using which their characteristic polynomials can easily be  obtained.

  From the above discussion it is clear that when $\h = \h (\mathbb C), \; T_g h_{m,n} = h^g_{m,n}$
  and $T_g$ leaves the $(L+1)$-dimensional subspace of $\h (\mathbb C)$ spanned by the vectors
\be
\mathfrak S(L) = \{ h_{L,0}, \; h_{L-1,1},\;  h_{L-2,2},\; \ldots , \; h_{0,L}\}
\label{rep-basis}
\en
invariant. Let $T(g,L)$ denote the restriction of $T_g$ to this subspace. Then the matrix elements of
$T(g,L)$ in the $\mathfrak S(L)$-basis are just the $M(g,L)_{rk}$ in (\ref{irrep-mat-elem}).

   There is an interesting intertwining relation between $M(g,L)$ and $T(g,L)$ \cite{ismail}
that is worth mentioning here. Note first, that using (\ref{comp-herm-poly9}) one can directly prove that
\be
  h_{m,n} (z, \overline{z}) = e^{-\partial_z\partial_{\overline{z}}}p_{m,n}(z, \overline{z}),
\label{int-op1}
\en
where
$$  p_{m,n}(z, \overline{z}) = \frac {z^m\overline{z}^n}{\sqrt{m!n!}} . $$
From this and the preceding discussion it is straightforward to verify that
\be
    e^{-\partial_z\partial_{\overline{z}}}M(g,L) = T(g,L)e^{-\partial_z\partial_{\overline{z}}}.
\label{int-op2}
\en

\section{Biorthogonal families of polynomials}\label{sec-biorth-poly}
From the way they were constructed in (\ref{full-basis}), it follows that the normalized complex Hermite polynomials $h_{m.n}\z$ form an orthonormal basis of $\h(\mathbb C)$
\be\label{e:ortho}
\int_{\C}\c{h_{m,n}\z} h_{k,l}\z \;d\nu\z = \delta_{mk}\;\delta_{nl} ,
\ee
where the bar denotes complex conjugation. Moreover,  two subspaces generated by  bases
$\mathfrak S(L)$ and $\mathfrak S(M)$, with $L \neq M$, are mutually orthogonal.
On the other hand the
$g$-deformed polynomials $h^g_{m,n}$ cannot be expected to form an  orthogonal set, except in very special cases. However, we have the following interesting result.

\begin{Theo}\label{theor-biorth}
The basis dual to $h_{m,n}\z, \;\; m + n = L, \;\; L =0, 1,2, \ldots ,$ in $\h(\mathbb C)$, consists of the deformed polynomials
\be
 \widetilde{h}^g_{m,n} = [T(g, L)^*]^{-1}h_{m,n} = h^{(g^*)^{-1}}_{m,n}, \qquad m+n = L.
 \label{dual-vecs}
 \en
which are biorthogonal with respect to the $h^g_{m,n}, \;\; m+n = L$, i.e.,
\be\label{e:biortho}
\int_{\C}\c{\widetilde{h}^g_{L-n, n} \z} h^g_{M-k,k}\z \;d\nu\z = \delta_{LM}\;\delta_{nk},
\ee
where, $n =0, 1,2, \ldots, L, \;\;  k =0, 1,2, \ldots, M$.
\end{Theo}

\prf
  Since the matrix $T(g,L)$, with matrix elements $M(g, L)_{rk}$, constitutes a representation of $GL(2,\mathbb C)$ on the subspace $\h_L(\mathbb C)$ of $\h(\mathbb C)$, generated by the basis $\mathfrak S(L)$, we know that $T(g,L)^{-1} = T(g^{-1},L)$. From (\ref{adj-rep}) it also follows that $T(g,L)^* = T(g^*,L)$. Thus the second equality in (\ref{e:biortho}) follows. An 
easy computation establishes the biorthogonality relation (\ref{e:biortho}).
 \qed
 
  \bigskip

   To summarize, the Hilbert space $\h(\C)$ decomposes into the orthogonal direct sum
$$  \h(\C) = \bigoplus_{L =0}^\infty \h_L (\C), $$
of $(L+1)$-dimensional subspaces $\h_L (\C)$, spanned by the orthonormal basis vectors $\mathfrak S (L)$, consisting of the complex Hermite polynomials $h_{L-k.k}, \; k =0,1,2, \ldots , L$. On each such subspace the operators $T(g,L), \; g \in GL(2,\C)$ define an
$(L+1)\times (L+1)$-matrix representation of $GL(2,\C)$. For each $g \in GL(2,\C)$ one obtains a set of $g$-deformed complex Hermite polynomials $h^g_{L-k.k} = T(g,L)h_{L-k,k}, \; k =0,1,2, \ldots , L$, in $\h_L (\C)$ and a biorthogonal set $\widetilde{h}^g_{L-k.k}, \; k =0,1,2, \ldots , L$, which constitute a family of
$g'$-deformed complex Hermite polynomials, with $g' = (g^{-1})^*$. In particular, when $g$ is the identity matrix, the two sets coincide with the (undeformed) complex Hermite polynomials $h_{m,n}$.

\section{Back to  noncommutative quantum mechanics}\label{sec-ncqm}
Let us specialize to hermitian group elements $g \in GL(2, \C)$ of the type
\be
  g = \begin{pmatrix} \alpha & \beta\\ \overline{\beta} & \alpha \end{pmatrix},
  \quad \alpha \in \mathbb R, \;\; 0 < \vert \alpha\vert < 1, \quad
  \beta = i\sqrt{1 - \alpha^2}\; .
\label{ncqm-mat}
\en
 For such a matrix we denote the deformed operators $a_i^g$ and ${a_i^g}^\dag$  by $a_i^\alpha$ and ${a_i^\alpha}^\dag$, respectively. They are seen to obey the commutation relations
\be
[a_i^\alpha, {a_i^\alpha}^\dag ] =1, \quad [a^\alpha_i, a^\alpha_j ] = 0, \quad [a^\alpha_1, {a^\alpha_2}^\dag ] = 2i \alpha\sqrt{1-\alpha^2} .
  \quad i,j = 1,2\; .
\label{ncqm-comm}
\en
These would be the same commutation relations as obeyed by the operators $A_i, \; A_i^\dag$ in (\ref{ceann-op-comm}) if we were to set $A_i = a_i^\alpha, \; A_i^\dag = {a_i^\alpha}^\dag$ and $\vartheta = 2 \alpha\sqrt{1-\alpha^2}$. In other words a matrix of the type (\ref{ncqm-mat}) is characteristic of a noncommutative model of quantum mechanics, obeying commutation relations of the type given in (\ref{nc-comm-relns1}) - (\ref{non-comm-relns2}) with $\gamma = \vartheta$. Denoting the associated deformed polynomials $h^g_{m,n}$ by $h^\alpha_{m,n}$
we say that the biorthogonal system of deformed complex Hermite polynomials $\{h^\alpha_{m,n}, \; \widetilde{h}^\alpha_{m,n}, \; m,n = 0,1,2, \ldots\}$, is naturally associated to this model of noncommutative quantum mechanics, in the same way as the orthonormal set of polynomials $h_{m,n}$ is associated to the standard quantum mechanics of two degrees of freedom. Note, however, that the matrix associated to $\widetilde{h}^\alpha_{m,n}$, which is the inverse of $g$ in (\ref{ncqm-mat}) is not of the same type as $g$. However, had we allowed a somewhat more general biorthogonality relation, e.g., of the type
$$
\int_{\C}\c{\widetilde{h}^\alpha_{L-n, n} \z} h^\alpha_{M-k,k}\z \;d\nu\z = \kappa_{L,n}\delta_{LM}\;\delta_{nk},
$$
where the $\kappa_{L,n}$ are positive constants, it would have been possible to find dual matrices under the action of which the commutation relations of noncommutative quantum mechanics would be preserved. As an example, consider the hermitian matrix
$$
  g' = \begin{pmatrix} \alpha & -\beta\\ -\overline{\beta} & \alpha
  \end{pmatrix}
$$
Comparing with (\ref{ncqm-mat}), we see that $g' g = \Delta\mathbb I_2$, where $\Delta$ is the determinant of $g$ (or $g'$) and $\mathbb I_2$ the
$2\times 2$ identity matrix. Defining the polynomials $\widetilde{h}^\alpha_{L-n, n}$ using this matrix, it is not hard to see from (\ref{irrep-mat-elem}) that $T(g', L)T(g, L) = \Delta^L\mathbb I_{L+1}$, so that $\kappa_{L,n} =\Delta^L$.

  To end this section let us note that we discussed here the model of non-linear
quantum mechanics in which we took $\vartheta = \gamma$, which means that the noncommutativity in the two position and the two momentum operators in (\ref{nc-comm-relns10}) and (\ref{non-comm-relns2}) are of the same amount. This meant that we had a system of two independent bosons as described by the commutation relations ((\ref{ceann-op-comm}), in particular the relation $[A_1, A_2] = 0$. On the other hand this condition was necessary to ensure that the ground state $\vert 0,0\rangle$ remained the same after the transformation (\ref{e:lin}).

\section{A note on some associated algebras of bilinear generators}\label{sec-bilin-ops}
We compile in this section some interesting results on algebras built out of bilinear combinations of the creation and annihilation operators ${a_i^\alpha}^\dag, \; a_i^\alpha, \; i = 1,2$, for a
fixed matrix of the type (\ref{ncqm-mat}). Note first that the operators
\be
J_1 = \frac 12 \Big( a_1^\dag a_2 + a_2^\dag a_1\Big), \quad
J_2 = \frac 1{2i} \Big( a_1^\dag a_2 - a_2^\dag a_1\Big), \quad
J_3 = \frac 1{2} \Big( a_1^\dag a_1 - a_2^\dag a_2\Big), \label{rot-gen}
\en
obey the commutation relations of the generators of $SU(2)$, i.e., $[J_1, J_2] = iJ_3$ (cyclic). If we add to this set $J_4 = \frac{1}{2}\Big(a^\dag_1 a_1 + a^\dag_2a_2\Big)$, we see that it commutes with the other three.
We would like to study the group(s) generated by the {\em deformed} operators
\be
\ja{1} = \frac{1}{2}\Big( \an{1}\ao{2} + \an{2}\ao{1}\Big), \;\;
\ja{2} = \frac{1}{2i}\Big( \an{1}\ao{2} - \an{2}\ao{1}\Big), \;\;
\ja{3} =  \frac{1}{2}\Big( \an{1}\ao{1} - \an{2}\ao{2}\Big),
\label{def-sec-ord-op}
\en
built by replacing the $a_i^\dag, \; a_i$ by the ${a_i^\alpha}^\dag, \; a_i^\alpha$,
which obey (see also \cite{kang}) the commutation relations
\be\label{e:genbo}
[\ao{j},\an{k}]= \delta_{jk} + \varepsilon_{jk}2i\alpha\sqrt{1-\alpha^2},
\ee
where $\varepsilon_{jk}$ is the usual anti-symmetric two-tensor. Using  (\ref{e:genbo}), we
get
\begin{align*}\label{e:var1}
[\ja{1}, \ja{2}] &= i \ja{3}, &  [\ja{2}, \ja{3}] &= i \ja{1},\\
[\ja{3}, \ja{4}] &= i\vartheta \ja{1}, & [\ja{4}, \ja{1}] &=  i\vartheta \ja{3}, \\
[\ja{3}, \ja{1}] &= i\ja{2} + i\vartheta \ja{4}, & [\ja{2}, \ja{4}] &= 0,
\addtocounter{equation}{1}\tag{\theequation}
\end{align*}
where again $\vartheta = 2\alpha\sqrt{1 - \alpha^2}$ and
$\ja{4}$ is defined as
\be
\ja{4} = \frac{1}{2}\Big(\an{1}\ao{1} + \an{2}\ao{2}\Big).
\ee
The above commutation relations are taken to hold for $0<\vartheta<1$ and $0 < \alpha < 1$.

 In order to analyze the Lie  algebra generated by the commutation relations
(\ref{e:var1}), which we denote by $\mathfrak g$, it is convenient to make a basis change.  We identify two
mutually commuting subalgebras $\fr{g}_{1}$ and $\fr{g}_{2}$:
\begin{align*}
\{X^{\vartheta}_{1}, X^{\vartheta}_{2}, X^{\vartheta}_{3}\} &\equiv \{ i\ja{1}, i\ja{3}, i(\ja{2} + \vartheta\ja{4})\} \in \fr{g}_{1}, \\
\{ Y^{\vartheta}\}& \equiv \{ \vartheta\ja{2} + \ja{4}\} \in\fr{g}_{2}.
\end{align*}
We then have the commutation relations,
\begin{align*}\label{e:lie2}
[X^{\vartheta}_{1},X^{\vartheta}_{2}] = X^{\vartheta}_{3}, \qquad [X^{\vartheta}_{2},X^{\vartheta}_{3}] &= (1-\vartheta^{2}) X^{\vartheta}_{1}, \qquad [X^{\vartheta}_{3},X^{\vartheta}_{1}] = (1-\vartheta^{2}) X^{\vartheta}_{2}, \\
&[Y^{\vartheta},X^{\vartheta}_{i}] = 0,
\addtocounter{equation}{1}\tag{\theequation}
\end{align*}
so that $\mathfrak g = \mathfrak g_1 \oplus \mathfrak g_2$.

In the limit $\vartheta \rightarrow 0$, (\ref{e:lie2}) leads to
\be\label{e:lie3}
[X^{0}_{i},X^{0}_{j}] = \varepsilon_{ijk}X^0_{k}, \quad [Y^0, X^0_i] = 0, \quad i,j = 1,2,3.
\en
In other words, $\fr{g}_{1}=\fr{su}(2)$ as expected and hence,
\be
\mathfrak g = \fr{su}(2)\oplus \fr{u}(1), \qquad \vartheta = 0.
\en

In the other limit, i.e., $\vartheta \rightarrow 1$ (the maximum value) we get
\be\label{e:lie4}
[X^{1}_{1},X^{1}_{3}] = [X^{1}_{2},X^{1}_{3}] = 0, \quad [X^{1}_{1},X^{1}_{2}] = X^{1}_{3},
\quad [Y^{1}, X^{1}_i] = 0,
\en
$i = 1,2,3$, which is a  nilradical basis, isomorphic to the Heisenberg algebra $\fr{h}$~\cite{hall}. Thus,
\be
  \mathfrak g = \mathfrak h \oplus \fr{u}(1), \qquad \vartheta = 1.
\en

Finally for $0 < \vartheta < 1$, with the re-scaled generators,
\be
  Z_1^\vartheta = \sqrt{1- \vartheta^2} X_1^\vartheta, \quad Z_2^\vartheta =X_2^\vartheta,
  \quad Z_3^\vartheta = \sqrt{1- \vartheta^2} X_3^\vartheta,
\label{srescaling}
\en
we again get
\be
  [Z_i^\vartheta, Z_j^\vartheta] = \varepsilon_{ijk}Z_k^\vartheta, \quad
  [Y^\vartheta , Z_i^\vartheta] =0, \quad i,j = 1,2,3,
\en
as in (\ref{e:lie3}). Once again,
\be
  \mathfrak g = \mathfrak{su}(2) \oplus \fr{u}(1), \qquad 0 < \vartheta < 1.
\en
 It is interesting note that, except in the case where $\vartheta = 1$, the algebra generated by the deformed generators is that of $\mathfrak{su}(2) \oplus \mathfrak{u}(1)$, exactly as in the undeformed case (\ref{rot-gen}). In the limit of $\vartheta = 1, \;\; \alpha^2 = \frac 12$
 and the commutation relation
 $[a^\alpha_1, {a^\alpha_2}^\dag ] = 2i \alpha\sqrt{1-\alpha^2}$  in (\ref{ncqm-comm}) becomes
$[a^\alpha_1, {a^\alpha_2}^\dag ] =  i$.

\section*{Acknowledgements}
 We would like to thank P. Winternitz, CRM, Universit\'e de Montr\'eal, for useful discussions. While working on this paper NMS would like to recognize the Ministry of Higher Education (MOHE) of Malaysia for the support she received under the Skim Latihan Akademik Institut Pengajian Tinggi (SLAI) Universiti Putra Malaysia (UPM) scholarship. One of us (STA) would like to acknowledge a grant from the Natural Science and Engineering Research Council (NSERC) of Canada.

\end{document}